\documentstyle[eqsecnum,aps]{revtex}

\def\ds{\d_{\! s}}
\def\nord{\raisebox{0.2ex}{$ \buildrel 
            {\hbox{\large .}}\over{\hbox{\large .}} $} }
\def\pparb{\cp_B^{
    \raisebox{.2ex}{${\scriptscriptstyle \,\parallel}$}} } 
\def\Jmrho{J_\m^{ {\scriptscriptstyle (} \r {\scriptscriptstyle )} } }

\def\dt#1{{\buildrel {\hbox{\large .}} \over {#1}}}  
\def\ointn{\oint d\s_\m}

\def\a{\alpha}
\def\b{\beta}
\def\c{\chi}
\def\d{\delta}
\def\g{\gamma}

\def\j{\psi}

\def\l{\lambda}
\def\m{\mu}
\def\n{\nu}
\def\o{\omega}
\def\p{\pi}                     
\def\r{\rho}                    
\def\s{\sigma}                  

\def\z{\zeta}

\def\G{\Gamma}

\def\O{\Omega}
\def\P{\Pi}

\def\S{\Sigma}

\def\cd{{\cal D}}

\def\cl{{\cal L}}
\def\co{{\cal O}}
\def\cp{{\cal P}}
\def\cs{{\cal S}}
\def\ct{{\cal T}}

\def\bfm#1{\mbox{\boldmath $#1$}}
\def\vev#1{\Big\langle #1 \Big\rangle}           
\def\svev#1{\left\langle #1\right\rangle}       

\def\beq{\begin{equation}}
\def\eeq{\end{equation}}
\def\bqry{\begin{eqnarray}}
\def\eqry{\end{eqnarray}}
\def\NON{\nonumber\\}
\def\seeq#1{eq.~(\ref{#1})}
\def\seEq#1{Eq.~(\ref{#1})}
\def\seeqs#1{eqs.~(\ref{#1})}

\def\seneq#1{~(\ref{#1})}

\def\JMP#1{Jour. Math. Phys. {\bf #1}}
\def\NPB#1{Nucl. Phys. {\bf B#1}}

\def\PLB#1{Phys. Lett. {\bf B#1}}
\def\PRD#1{Phys. Rev. {\bf D#1}}

\def\PRP#1{Phys. Rep. {\bf #1}}

\begin{document}
\draft
\preprint{TAUP--xxxx--96}
\twocolumn[\hsize\textwidth\columnwidth\hsize\csname%
@twocolumnfalse\endcsname

\title{
$
\phantom{aaaaaaaaaaaaaaaaaaaaaaaaaaaaaaaaaaaaaaaaaaaaaaaaaaaaaaaaaaaaaaaaaaaaa}$
{\rm TAUP--2399--96}\\A Supersymmetry Anomaly}
\author{Aharon Casher and Yigal Shamir\cite{supp}}
\address{ School of Physics and Astronomy,
Beverly and Raymond Sackler Faculty of Exact Sciences\\
Tel-Aviv University, Ramat Aviv 69978, ISRAEL}
\date{\today}
\maketitle

\begin{abstract}
A supersymmetry anomaly is found in the presence of non-perturbative fields.
When the {\it action} is expressed in terms of the 
correct quantum variables, anomalous surface terms appear in its 
supersymmetric variation -- one per each collective coordinate. 
The anomalous surface terms do not vanish in general when inserted 
in two- or higher-loop bubble diagrams, and generate a violation of the SUSY
Ward identities.
\end{abstract}
\pacs{email: {\it ronyc@post.tau.ac.il \hspace{2ex} 
shamir@post.tau.ac.il}}
]
\narrowtext
\renewcommand{\theequation}{\arabic{equation}}

\noindent {\it 1.} 
The  assumption that supersymmetry (SUSY) is not broken {\it explicitly} by
non-perturbative  quantum effects underlies many investigations of
supersymmetric gauge theories~\cite{conv,sw}. 
This assumption was tested in numerous semi-classical instanton calculations, 
but its validity beyond the semi-classical 
approximation has remained elusive~\cite{mrs,cs}.

  No consistent UV regularization known today preserves SUSY. 
Also, one cannot define a SUSY theory in a finite box, 
because the SUSY variation involves the canonical fields as well as 
their conjugate momenta, and no choice of boundary conditions can ensure 
a vanishing SUSY current at the boundaries. These (not unrelated) facts 
mean that  the possibility that quantum effects 
do violate SUSY merits a careful investigation.

  In perturbation theory, conservation of the SUSY current can be
enforced order by order. However, the significance of this
observation is limited, since perturbation theory
is an asymptotic expansion. Whether or not SUSY is an exact symmetry, 
is a question that must be settled by going
beyond perturbation theory.

When spontaneous SUSY breaking does not take place, 
SUSY should be realized via exact Ward identities (WI) of the 
general form $\svev{\ds\co}=0$. Here $\co$ is a 
gauge invariant (multi)local operator, and $\ds$ stands for the SUSY
variation. In order to examine the validity of SUSY WIs, we have calculated 
$\svev{\ds\co}$ in the most general continuum path-integral framework, 
using the standard rules of Quantum Field Theory~\cite{thft}. 
The result is the following equation
\beq
  \vev{\ds\co} = \sum_n \vev{\co\, \ds\z_n\, \ointn \nord J^n_\m \nord } \,.
\label{an}
\eeq
The sum in \seeq{an} runs over the collective coordinates pertaining to
a given non-perturbative sector. $\ds\z_n$ is
the SUSY variation of the $n$-th collective coordinate. 
The current $J_\m^n$ involves a\ $\z_n$-derivative of the quantum
field, and its explicit form is given below. The surface integration
is at space-time infinity. The normal-ordering symbol
means that the one-loop diagram obtained by contracting  
the two fields in $J_\m^n$ is discarded. 
  
  The SUSY anomaly is a sub-leading effect
that arises from the interplay between two different physical scales.
Previous non-perturbative calculations were restricted to the semi-classical 
approximation~\cite{conv}, which is too crude to expose any anomalous breaking 
of SUSY. An exception is the two-loop result of ref.~\cite{mrs}, 
which, however, has a limited scope, because it pertains to Super Yang-Mills, 
a theory that does not admit a systematic small-$g$ expansion due to 
IR divergences. In contrast, our results are directly applicable to any 
non-perturbative sector of a weakly-coupled SUSY theory, 
and to any desired order. 

  In an operator language, we find a violation of the SUSY
algebra at the non-perturbative level. A similar pattern exists in
the case of the chiral anomaly -- the axial charge $Q_5$ is conserved
to all orders in perturbation theory, and only when non-perturbative
effects are taken into account does one have $\dt{Q}_5 \ne 0$.

  The non-conservation of $Q_5$ is a consequence of the existence of 
zero modes in the eigenmode expansion of the fermion fields. 
Neither the UV regularization, nor the triangle graph,
play a direct role. While the details
are different, and more complicated, in the SUSY case, it remains true that the 
UV regularization plays little role in what follows.
  
  \seEq{an} stems from a feature of the {\it action principle}, 
whose significance has not been  appreciated before -- {\it the surface terms
in the variation of the action depend in a non-trivial way 
on the nature of the independent field variables}. Specifically,
when the action is expressed in terms of the quantum variables
pertaining to a non-perturbative sector, its SUSY variation contains  
the anomalous surface terms that ultimately appear in \seeq{an}.

\vspace{2ex}
\noindent {\it 2.}
We first specify our notation. In a non-perturbative sector, 
each bosonic field $B(x)$ is split into a classical and a quantum part 
$B(x) = b(x) + \b(x)$. Here $b(x)$ is the classical field
and $\b(x)$ is the quantum part. It is {\it not} assumed that $b(x)$ is
an exact solution of the classical field equations~\cite{thft}. 
Fermion fields are denoted $\j(x)$.
Also, $Q(x)$ denotes {\it any} field, 
and $\hat{q}(x)$ is its quantum part. 
(If $Q(x)=\j(x)$ is a fermion field, then $Q(x)=\hat{q}(x)$.)
The background field $b(x)$ depends explicitly on the 
collective coordinates $\z_n$. In addition, an infinite number of
{\it gauge degrees of freedom} $\o^a(x)$ should be fixed.

 The quantum part of each field is expanded in terms of independent 
modes $\hat{q}(x) = \int' dp\, \c_p(x) \hat{q}_p$.
The amplitude of a quantum mode is 
denoted $\hat{q}_p$. The corresponding eigenfunction is $\c_p(x)$,
with eigenvalue $\l_p$.
The symbol $\int' dp$ stands for $\sum \int (dp/2\p)$ where 
$p=\sqrt{p_\m p_\m}\,$ and $\sum$ is over all other quantum numbers
of the continuous spectrum, plus a sum over normalizable states.
In particular, fermionic zero modes
are included in the eigenmode expansion of the fermion fields.
The quantum part of the bosonic fields satisfies $\pparb\, \b = 0$, where
\beq
   \pparb = \sum_{IJ} \bfm{|} b_{;I} \bfm{)}\bfm{(} b_{;I} \bfm{|} b_{;J} \bfm{)}^{-1} \bfm{(} b_{;J} \bfm{|} \,,
\label{prll}
\eeq 
and  $I=(n,\o)$ is a generic index. 
$b_{;\o}(y) \equiv \O_a(x)\d(x-y)$, where $\O_a(x)$ is the 
linear differential operator such that 
$\O\,\o(x)$ is an infinitesimal 
local gauge transform of the classical field. 
$b_{;n}(y)=b_{,n}+\O\,\o_n$ where $\o_n$ is determined by the background gauge 
condition $\O^\dagger b_{;n}=0$.
The bosonic eigenfunctions are determined by the 
eigenvalue equation $L_B^\perp\, \c^B = \l^2 \c^B$. 
Here $\l^2$ is an eigenvalue of $p^2 + M^2$ 
where $M$ is the mass matrix.
$L_B^\perp = \cp_B^\perp L_B \cp_B^\perp$ where $L_B$ is the
differential operator that enters the quadratic part of the bosonic action,
and $\cp_B^\perp(x,y) = \d(x-y) - \pparb(x,y)$ is a transversal projector.
Notice that all eigenfunctions and, hence, all quantum fields $\hat{q}(x)$,
are {\it explicit} functions of the collective coordinates $\z_n$.

  Consider an infinitesimal field transformation $Q(x)\to Q(x)+\d Q(x)$.
To keep the discussion general, we only assume at the moment that 
$\d Q(x)$ is a local function of $Q(x)$, $Q_{,\m}(x)$, etc. 
In a functional integral, {\it the field variation $\d Q(x)$ must be generated 
by variations of the independent variables}, which include 
the collective coordinates $\z_n$, the gauge degrees of freedom $\o^a(x)$, 
and the amplitudes of the quantum modes $\hat{q}_p$. Explicitly,
\beq
\d Q(x) = \d \hat{q}(x) + \sum_n Q_{;n}(x) \d\z_n + Q_{;\o}(x) \d\o(x) \,.
\label{var}
\eeq
Here $\d \hat{q}(x) \equiv \int' dp\, \c_p(x) \d\hat{q}_p$.
\seEq{var} {\it defines} the independent variations $\d\z_n$, $\d\o^a(x)$
and $\d\hat{q}_p$
in terms of $\d Q(x)$. The field derivative $Q_{;I}$ is given by
$b_{;I} + \b_{;I}$ for bosons ($\j_{;I}$ for fermions). 
For the quantum part of any field
$\hat{q}_{;\o}(x) = -igT^a\hat{q}(x)$
and $\hat{q}_{;n}(x) = \int' dp\, \c_{p;n}(x) \hat{q}_p$
where $\c_{p;n} = \c_{p,n} - igT^a \o^a_n \c_p$.
An explicit expression for $\d\z_I=(\d\z_n,\d\o^a(x))$ 
follows from \seeq{var} by exploiting the  
constraint $\pparb\b=0$. In the SUSY case one has
$\ds\z_I = \sum_J C^{-1}_{IJ}\, \bfm{(} b_{;J} \bfm{|} \G \j \bfm{)}$.
In this expression, $\G$ is a constant matrix (that matches the indices 
of bosons and fermions), $\G\j=\ds B$ is the SUSY variation of bosons, and
$C_{IJ} = \bfm{(} b_{;I} \bfm{|} b_{;J} + \b_{;J} \bfm{)}$.

  We now turn to the derivation of \seeq{an}. 
We introduce the functional differentiation operator
\beq
  \ct =  \int' dp\, \d\hat{q}_p {\partial\over \partial\hat{q}_p}
         + \sum_n \d\z_n\, {\partial\over \partial\z_n} \,.
\label{trans}
\eeq
We will also write $\ct_q$ and $\ct_\z$ to denote  respectively
the first and second terms on the r.h.s.\ of \seeq{trans}.
For any local gauge-invariant operator $\co$, one has $\ct\co=\d\co$,
where by definition,
$\d\co \equiv \d Q (\partial\co/\partial Q) + 
\d Q_{,\m} (\partial\co/\partial Q_{,\m}) + \cdots$.
The expectation value $\svev{\d\co}$ is given by the functional integral
\beq
  \vev{\d\co} = \int \prod_n d\z_n\,  \cd \hat{q}\, J\, e^{-S}\, \ct\co \,,
\label{TO}
\eeq
where 
$J={\rm Det}\, C / 
{\rm Det}\, \bfm{(} b_{;I} \bfm{|} b_{;J} \bfm{)}^{1\over 2}$ 
is a generalized Fadeev-Popov jacobian~\cite{mrs}. 
$C_{IJ}$ is the same matrix that enters the definition of $\d\z_I$ above.
Integration by parts now leads to the fundamental WI
\beq
  \vev{\d\co} = \vev{\co\, (\ct S)} + \vev{\co\, \d \m} \,.
\label{WI}
\eeq
Here $\d\m$ is (minus) the variation of the path integral measure
\beq
  -\d\m \equiv \ct \log J 
  + \int' dp\, {\partial\d\hat{q}_p \over \partial\hat{q}_p} 
  + \sum_n {\partial\d\z_n \over \partial\z_n} \,.
\label{dm}
\eeq

  We comment in passing that the above reasoning is completely general.  
It is instructive to see how the formalism works in the case of an 
infinitesimal translation. Let us define $\d_\m Q = Q_{,\m}$. 
In this case, all collective and quantum variables are invariant,
except for the translation collective coordinates $x^0_\n$,
which transform according to $\d_\m x^0_\n = -\d_{\m\n}$. 
Applying \seeqs{WI} and\seneq{dm}, one easily verifies
that $\svev{\co_{,\m}}=0$.

  We now have to compute $\ct S$, paying attention to the surface terms.
It is convenient to start with the action principle
\bqry
  \int d^dx\, \d\cl & - & \ointn (\P_\m \d Q) = \NON
  & & \int d^dx\, \d Q \left( -\P_{\m,\m} + \partial\cl/\partial Q \right) \,.
\label{eom}
\eqry
As usual $\P_\m = \partial\cl/\partial{Q_{,\m}}$ \,where the lagrangian 
$\cl=\cl(Q,Q_{,\m})$.
Substituting \seeq{var} into the r.h.s. of \seeq{eom},
and integrating by parts leads to 
\setcounter{equation}{0}
\renewcommand{\theequation}{9\alph{equation}} 
\begin{eqnarray}
  \int d^dx\, & & \hspace{-0ex} 
  \d Q \left( -\P_{\m,\m} + \partial\cl/\partial Q \right) = 
\NON
  & & \int d^dx\, \d \hat{q} 
  \left( -\P_{\m,\m} +\partial\cl/\partial Q \right)
\label{tq} \\
  & & + \sum_n \d\z_n \int d^dx\,  \Big( \p_\m Q_{,\m,n} +
  (\partial\cl/\partial Q) Q_{,n} \Big) 
\label{tz} \\
  & & - \sum_n \d\z_n \ointn J_\m^n \,.
\label{intj}
\end{eqnarray}
\renewcommand{\theequation}{\arabic{equation}}
\setcounter{equation}{9}
\hspace{-2.5ex} We have used the gauge invariance of the action. Here
\beq
  J_\m^n = \P_\m Q_{,n} \,,
\label{J}
\eeq
In terms of the modes, the bilinear part of the action is
$S^{(2)} = (1/2) \int' dp\, \l_p \hat{q}_p^2$. Therefore,
\setcounter{equation}{0}
\renewcommand{\theequation}{11\alph{equation}} 
\begin{eqnarray}
  \ct_q S^{(2)} & = & \int' dp\,  \l_p \hat{q}_p \d\hat{q}_p 
\label{ds2} \\
  & = & \int d^dx\, \d\hat{q}(x) L \hat{q}(x) \,.
\label{ds2x}
\end{eqnarray}
\renewcommand{\theequation}{\arabic{equation}}
\setcounter{equation}{11}
\hspace{-2.5ex} In going from\seneq{ds2} to\seneq{ds2x},
we must let $L$ act on $\hat{q}(x)$, and not on $\d\hat{q}(x)$.
This implies that expression\seneq{tq} is equal to $\ct_q S$.
(Note that $\partial\cl/\partial Q = \partial\cl/\partial \hat{q}$.)

  Since $\partial/\partial\z_n$ acts on the classical field or on
the wave-functions, expression\seneq{tz} is equal to $\ct_\z S$. 
Putting everything together, we arrive at the following result
\beq
  \ct S  =  \int  d^dx\, \d\cl -  \ointn (\P_\m \d Q)
        + \sum_n \d\z_n \ointn J_\m^n \,.
\label{TS}
\eeq 
\seEq{TS} is completely general. 
In the SUSY case, $\ds\cl$ is a total derivative,
and the first two terms on the r.h.s.\ of \seeq{TS} are
equal to $\ointn S_\m$. 
{\it The last term in \seeq{TS} is an anomalous surface term}.
($\ointn S_\m$ vanishes unless there are 
massless one-fermion states in the spectrum of the SUSY current $S_\m$.)
In the above discussion, one can neglect modifications of the lagrangian
by a total derivative $K_{\m,\m}$. This results in adding
the $\z_n$-derivative of a current $K_{\m,n}$ to $J_\m^n
$, 
which leaves invariant our final expression for the anomaly
(see \seeq{srho} below).

  It remains to compute $\ds\m$. Using \seeq{var}, the expressions
for $\ds\z_n$ and $\ds\hat{q}_p$, and some lengthy algebra  we get
\beq 
  \vev{\ds\co} = \sum_n \vev{\co\, \ds\z_n \Big( {\rm STr}\, 
  (\partial/\partial\z_n) + \ointn J_\m^n \Big)} \,.
\label{an'}
\eeq 
The spectral trace 
${\rm STr}\, (\partial/\partial\z_n) = 
\int' dp\, (-)^F \bfm{(} \c_p \bfm{|} \c_{p,n} \bfm{)}$ 
comes from $\ds\m$. In a gauge theory it
includes the contribution of the ghost fields.

  Regardless of its precise definition, the spectral trace
{\it cannot cancel} $\ointn J_\m^n$, because the former is a $c$-number 
function of the collective coordinates, whereas the latter is an operator. 
In fact, a detailed diagrammatic analysis which we relegate to a 
separate publication~\cite{long}, 
shows that {\it subtracting} $-{\rm STr}\, (\partial/\partial\z_n)$ 
from $\ointn J_\m^n$ yields the  
normal-ordering prescription defined in \seeq{an}.  

In terms of diagrams (compare \seeq{TO}),
\beq
  \vev{\co} = \int \prod_n d\z_n\,\, e^{-S_{cl}-W_1-W_2 }\, 
  \vev{\co}_\z \,.
\label{vevz}
\eeq
At fixed $\z_n$, $\exp(-W_1)$ is 
${\rm Det}\, \bfm{(} b_{;I} \bfm{|} b_{;J} \bfm{)}^{1\over 2}$
times the functional determinants of $L_B^\perp$ and $L_F$. $W_2$ is the sum
of all connected bubble diagrams, and $\svev{\co}_\z$ 
is the sum of all diagrams that together constitute an insertion of $\co$. 
Using this notation,
the result of the diagrammatic calculation is 
\bqry
  \vev{\ds\co}_\z & = & \sum_n \left\{
  \vev{\co\, \ds\z_n \ointn \nord J_\m^n \nord }_\z 
  + {\partial \over \partial\z_n} \vev{\co\, \ds\z_n}_\z
  \right. \NON
  & & - \left. \vev{\co\, \ds\z_n}_\z \,\, {\partial \over \partial\z_n} 
  \left( S_{cl}+W_1+W_2 \right) \right\} \,.
\label{anz}
\eqry
\seEq{anz} is valid in the (background) Landau gauge. 
Multiplying \seeq{anz} by $\exp(-S_{cl}-W_1-W_2)$, integrating over
$\z_n$, and dropping a total $\z_n$-derivative, one arrives at \seeq{an}.
In the semi-classical approximation, this result is consistent with 
the instanton calculus of Novikov {\it et.\ al.}~\cite{conv}. 

  For an operator $\int d^dx\, \co(x)$,
which is the integral over all space-time of a local density,
\seeqs{an'} and\seneq{anz} are modified by an additional surface term.
This is true in particular for the SUSY generators. The additional
surface term represents an anomalous transformation law, and it arises 
for the same reasons as the anomalous surface term in the variation of the 
action. See ref.~\cite{long} for the details. (SUSY violations arising from the 
UV regularization are cancelled by tuning the counter-terms order by order.
This mechanism works both in the vacuum sector and in non-perturbative 
sectors, and is taken into account in \seeq{anz}.
The only remaining effect of the UV regulatization amounts to anomalous
{\it local} terms in the transformation law of certain composite
operators~\cite{k}.) 

\vspace{2ex}
\noindent {\it 3.} 
We now consider diagrams with an insertion
of $\ointn \nord J_\m^n \nord$.  Terms in $J_\m^n$ involving 
the classical field, as well as trilinear terms
(which arise in a gauge theory) vanish at infinity and do not
contribute to\seneq{intj}. (For the trilinear terms this is 
due to transversality of the bosonic quantum field.)
Therefore, henceforth 
$J_\m^n = \hat\p_\m \hat{q}_{,n}$,
where $\hat\p_\m = \hat{q}_{,\m}$ for bosons and ghosts, 
and $\hat\p_\m = \bar\j \g_\m$ for fermions.

  For an asymptotically large distance $R$, the matrix element of $J_\m^n$
between eigenstates of momenta $p$ and $p'$
involves an oscillatory factor $\exp(\pm iR(p \pm p'))$. 
{\it A non-vanishing result is therefore  possible only when
$\ointn \nord J_\m^n \nord$ is inserted into a bubble diagram},
because the latter contains a piece proportional to \mbox{$\d(p-p')$}. 
The anomalous term in \seeq{anz} thus factorizes into
$\svev{\co\, \ds\z_n}_\z \, \svev{\ointn \nord J_\m^n \nord }_\z$.
 
A diagram with an insertion of $\ointn \nord J_\m^n \nord$ contains one 
special line $G_n(x,y) = \svev{\hat{q}(x) \hat{q}_{,n}(y)}$.
Explicitly
\beq
  G_n(x,y) = \int' dp\, \c_p(x) {1\over \l_p} \c^*_{p,n}(y) \,.
\label{Gn}
\eeq
All other lines in the diagram correspond to ordinary propagators 
$G(x,y)$ in the given non-perturbative sector.
However, since the surface integration takes place at infinity, one can replace
$G(x,y)$ by the corresponding free propagator. Observe that in general 
(except for collective coordinates related to exact symmetries
such as global translations)
$G_n(x,y)$ is {\it not} a linear combination of ordinary propagators and 
their derivatives even asymptotically.

  As a representative example, we now write down
an explicit expression for the sum of all bubble diagrams
with an insertion of $\ointn \nord \Jmrho \nord$, 
where $\r$ is the scale collective coordinate of the one-instanton sector.
Exploiting the spherical symmetry
of the instanton field, and removing kinematical factors, the
asymptotic behaviour of a {\it out}-state is
$\c_p \sim \exp(-ipr) + \cs \exp(ipr+i\a_l)$, where $\a_l$ is a 
constant phase that depends only on the angular momentum $l$. 
$\cs=\cs(p;l;\r)$ is the first-quantized S-matrix associated with
the differential operator $L^\perp_B$ for bosons, with
$L_F L_F^\dagger$ for fermions (we consider vector-like fermions
for simplicity), and with $\O^\dagger \O$ for ghosts. At fixed $\r$,
\bqry
  \vev{\ointn \nord \Jmrho \nord}_\r =
  - \, {\rm tr}\, (-)^F \! 
  \int {dpdp'\over 8\p^2}\,\, \tilde\S(p') \tilde{G}(p') 
  \hspace{2ex}
\NON 
   \times \left[
  \ointn \Jmrho(p',p) \, G_0(p) \, \bfm{(} \c_p \bfm{|} \c_{p'}^0 \bfm{)} \, 
   + {\rm h.c.} \right] .
\label{sclr}
\eqry
$\tilde{G}^{-1}(p)=G_0^{-1}(p)+\tilde\S(p)$, where $G_0(p)$
is the free propagator and $\tilde\S(p)$ 
is the self-energy in the {\it vacuum sector}. $\Jmrho(p',p)$
is the matrix element of $\Jmrho$ between $\c_{p'}^0$ and $\c_p$, where
$\c_{p'}^0$ is a free-field eigenstate. Keeping only the singular part,
$\bfm{(} \c_p \bfm{|} \c_{p'}^0 \bfm{)} \approx \p\d(p-p')(I+\cs^\dagger)$ 
while 
$\tilde{G}_0(p)\ointn \Jmrho(p,p) = ip \l^{-2}\, \partial\cs/\partial\r$. 
(Recall that $\l^2$ diagonalizes $p^2+M^2$.) We thus arrive at
\bqry
  {\vev{\ointn \nord \Jmrho \nord}}_\r = 
  -i\, {\rm tr}\, (-)^F \! 
  \int {d\l\over 8\p\l}\, \tilde\S \tilde{G} \hspace{2ex}
\NON 
  \times [(\partial \cs / \partial\r)(I+\cs^\dagger) - {\rm h.c.}] \,.
\label{srho}
\eqry
This result generalizes to background fields lacking a spherical symmetry, 
and constitutes an explicit expression for the anomaly.
Bubble diagrams with an insertion of $\ointn \nord J_\m^n \nord$
have no overall UV divergence, because $\partial \cs / \partial\z_n$ always 
vanishes rapidly for large $p$. 

  An {\it algebraic} relation between the small-fluctuations operators 
follows from \seeq{eom}. It reads
\beq
\G L_B - L_F L_F^\dagger \G \:\: {\buildrel \circ \over = } \:\:
  V \cdot \ds\j(b) \,.
  \label{bf}
\eeq
The symbol $\, {\buildrel \circ \over = } \,$  means that \seeq{bf}
holds when applied to functions $\c^B(x)$ obeying the background gauge 
condition $\O^\dagger \c^B = 0$.
$\ds\j(b)$ is the classical part of the SUSY variation of 
the fermions, $\G$ has been defined below \seeq{var}, and
$V$ is the three-index coupling defined by the fermion interaction lagrangian
$\cl_F^{int} = V_{AbC} \j_A \b_b \j_B$. 

  \seEq{bf} implies that the scattering potentials that enter 
the bosonic and fermionic small-fluctuations equations are different.
Consequently, $\cs_B \ne \cs_F$.
(In four dimensions, the only exceptions are pure YM 
instantons with no Higgs field, 
where, due to the self-duality of {\it exact} solutions, 
the r.h.s.\ of \seeq{bf} involves a Dirac projection operator
on a single chirality.) Since $\cs_B \ne \cs_F$, 
and since the all-orders expansion of $\tilde\S \tilde{G}$
constitutes an infinite set of linearly independent functions of $p$,
the sum of all bubble diagrams
${\svev{\ointn \nord \Jmrho \nord}}_\r$ cannot vanish. This,
in turn, implies that ${\svev{\ds\co}}$ is non-zero in general,
because $\co$ and the $\r$-dependence of $\svev{\co\, \ds\r}_\r$ are arbitrary.

  In practice, ${\svev{\ointn \nord \Jmrho \nord}}_\r$ 
should be $O(g^4)$. First, $\tilde\S$ is manifestly $O(g^2)$.
An additional $O(g^2)$ suppression factor appears because the fluctuations
spectrum is approximately supersymmetric in the instanton's core.
The $\r$-integration is dominated by $\r\sim v^{-1}$, 
where $v$ is the Higgs VEV~\cite{thft}. On the other hand, effects of the
position-dependent background Higgs field, 
which are responsible for the discrepancy between 
$\cs_B$ and $\cs_F$, arise from the mass scale $m=gv$. As a result, the
relevant form-factors depend on the dimensionless variable $m\r \sim m/v =g$.

\vspace{2ex}
\noindent {\it 4.}
In this Letter we have derived a general expression for the SUSY anomaly.
Our results prove that SUSY violations arise generically
in asymptotically-free and super-renormalizable theories, 
when the full non-perturbative effects of 
the physical IR cutoff are taken into account. The essential
ingredients of the violation are the existence of collective
coordinates which transform non-trivially under SUSY, and the
discrepancy between the dependence 
of the first-quantized S-matrices of bosons and fermions
on the collective coordinates.

  We now summarize the situation is various euclidean dimensions.
In one dimension (SUSY q.m.) $\cs_B$ is always equal to $\cs_F$
on the relevant backgrounds. For two- and three-dimensional single instantons,
{\it i.e.} fluxons and monopoles respectively, 
$\cs_B \ne \cs_F$, but both do not depend on the 
collective coordinates, which correspond only to exact symmetries.
The single four-dimensional instanton was discussed above.
Finally, multi instanton-antiinstanton configurations in more than one 
dimension lead to a non-trivial dependence of $\cs_B -\cs_F$ on {\it some}
collective coordinates (such as relative positions).

{\it Acknowledgement}: We thank N.\ Itzhaki for his comments on the 
manuscript.

\end{document}